\documentclass[aip,jcp,amsmath,amssymb,floatfix,reprint,citeautoscript]{revtex4-2}

\usepackage[utf8]{inputenc}
\usepackage{graphicx}
\usepackage{wrapfig}
\usepackage{placeins}
\usepackage[colorlinks,allcolors=black,citecolor=blue,urlcolor=blue]{hyperref}
\usepackage[version=3]{mhchem}
\usepackage{chemformula}
\usepackage{siunitx}

\usepackage{float}

\graphicspath{{figures/}}

\begin{document}
\newcommand{\icm}{\,cm\textsuperscript{-1}}

\def\mytitle{
Infrared spectra at coupled cluster accuracy from neural network representations
}
\title{\mytitle}

\author{Richard Beckmann}
\email{richard.beckmann@rub.de}
\affiliation{%
Lehrstuhl f\"ur Theoretische Chemie,
Ruhr-Universit\"at Bochum, 44780 Bochum, Germany
}
\author{Fabien Brieuc}
\affiliation{%
Lehrstuhl f\"ur Theoretische Chemie,
Ruhr-Universit\"at Bochum, 44780 Bochum, Germany
}
\affiliation{%
Current Address: Laboratoire Mati\`ere en Conditions Extr\^emes, 
Universit\'e Paris-Saclay, CEA, DAM, DIF, 91297 Arpajon, France
}

\author{Christoph Schran}%
\email{christoph.schran@rub.de}
\affiliation{%
Lehrstuhl f\"ur Theoretische Chemie,
Ruhr-Universit\"at Bochum, 44780 Bochum, Germany
}
\affiliation{%
Current Address: Yusuf Hamied Department of Chemistry, University of Cambridge, Lensfield Road, Cambridge, CB2 1EW, UK
}
\author{Dominik Marx}
\affiliation{%
Lehrstuhl f\"ur Theoretische Chemie,
Ruhr-Universit\"at Bochum, 44780 Bochum, Germany
}

\date{\today}

\begin{abstract}
Infrared spectroscopy is key to elucidate 
molecular structures, monitor reactions and observe conformational changes, 
while providing information on both structural and dynamical properties.
This makes the 
accurate
prediction of infrared spectra based on first-principle theories
a highly desirable pursuit.
Molecular dynamics simulations have proven to be a particularly powerful approach 
for this task, albeit  requiring the computation
of energies, forces and dipole moments for a large number of molecular configurations
as a function of time.
This explains why highly accurate first principles methods, such as coupled cluster theory, 
have so far been inapplicable for the prediction of 
fully anharmonic vibrational spectra of large systems at finite temperatures. 
%
Here, we push cutting-edge machine learning techniques forward
by using neural network representations 
of energies, forces and in particular dipoles 
to predict such infrared spectra 
fully 
at 
``gold standard''
coupled cluster accuracy as demonstrated for
protonated water clusters 
as large as
the protonated water hexamer, 
in its extended Zundel configuration.
Furthermore, we show that this methodology can be used beyond the
scope
of the data considered during the development of the 
neural network models, 
allowing for the computation of finite-temperature infrared spectra of
large systems
inaccessible to explicit coupled cluster calculations.
This substantially expands the 
hitherto existing 
limits of accuracy, speed and system size
for theoretical spectroscopy and opens up a multitude of
avenues for the prediction of vibrational spectra
and the understanding of complex intra- and intermolecular couplings.
\end{abstract}

{\maketitle}

\section{Introduction}
\label{sec:intro}

Infrared~(IR) spectroscopy is one of the most 
useful analytical tools in chemistry
to identify compounds and understand their properties.
This is exemplified in studies on the detection of carbon-rich building blocks in space~\cite{McGuire_10.1126/science.abb7535},
the nano-imaging of materials~\cite{Huth_10.1038/nmat3006},
the observation of dynamical couplings in peptides~\cite{Kolano_10.1038/nature05352},
or the imaging of the proton transfer mechanism in water~\cite{Wolke_10.1126/science.aaf8425,Thamer_10.1126/science.aab3908,Dahms_10.1126/science.aan5144}
to name but a few examples.
However, the interpretation of the measured IR spectra
becomes increasingly complicated when moving to more complex systems.
Theoretical IR spectroscopy from molecular dynamics
(MD)
with classical or quantum nuclei~\cite{Ivanov_10.1039/C3CP44523B} 
provides a unique route to unravel the details of the observed spectra,
but requires the highest accuracy in the description
of the electronic structure in order to be predictive.
Such accuracy is nowadays in many cases best provided by coupled cluster~(CC) theory~\cite{Bartlett_10.1103/RevModPhys.79.291,Hobza_10.1021/ar200255p} ---
the current ``gold standard'' in quantum chemistry.
At the same time, the 
computational cost of 
CC~methods,
which remains enormous despite much progress in linear-scaling techniques\cite{Riplinger_10.1063/1.4821834,Liakos_10.1021/acs.jctc.5b00359}, 
prevents their routine usage for the simulation of IR~spectra~--
except for small prototypical systems and in the absence of temperature
and nuclear quantum effects. 
%

Within the last decade, the rise of machine learning~(ML)
in chemical physics~\cite{Behler_10.1063/1.4966192,
Butler_10.1038/s41586-018-0337-2,
Deringer_10.1002/adma.201902765
}
has created the opportunity to represent the potential energy surface~(PES), 
governing the dynamics of a given system, 
at substantially reduced computational cost.
With high-dimensional neural network potentials~(NNPs) paving the way\cite{Behler_10.1103/PhysRevLett.98.146401,
Behler_10.1002/anie.201703114,
Behler_10.1021/acs.chemrev.0c00868}, 
a multitude of different techniques has been developed to create
highly accurate models of
interactions~\cite{
Ghasemi_10.1103/PhysRevB.92.045131,
Schuett_10.1038/ncomms13890,
Zhang_10.1103/PhysRevLett.120.143001,
Bartok_10.1103/PhysRevLett.104.136403, 
Rupp_10.1103/PhysRevLett.108.058301,
Thompson_10.1016/j.jcp.2014.12.018,
Shapeev_10.1137/15M1054183,
Li_10.1103/PhysRevLett.114.096405,
Chmiela_10.1126/sciadv.1603015
};
see in particular Ref.~\citenum{Manzhos_10.1021/acs.chemrev.0c00665}
for a detailed review with a focus on small molecules and reactions.
In recent years, ML~approaches have
progressed towards the description of properties,
such as 
polarizabilities~\cite{Wilkins2019/10.1073/pnas.1816132116} or
dipole moments~\cite{Gastegger_10.1039/c7sc02267k,%
Peyton2020/10.1021/acs.jpca.0c02804} which modulate the 
IR~spectral intensities.
This has been first shown for 
electric dipole moments in Ref.~\citenum{Gastegger_10.1039/c7sc02267k} based on
environment-dependent charges represented by neural networks.
Later work has further improved the accuracy of ML~dipole models by a combination 
of atomic dipole moments and charges~\cite{Veit_10.1063/5.0009106,
Grisafi_10.1103/PhysRevLett.120.036002
},
or extended the formalism to transition dipoles for the
prediction of UV absorption spectra~\cite{Westermayr2020/10.1063/5.0021915}.
In recent work, the simultaneous prediction of energies, forces and dipole moments 
has been realized in approaches like 
PhysNet~\cite{Unke_10.1021/acs.jctc.9b00181}
and Schnetpack~\cite{Schnetpack}.
%
Very elegantly, the dipole moments can be incorporated 
into the ML model as the response
of the energy model to an external electric field\cite{Christensen_10.1063/1.5053562,Gastegger_10.1039/D1SC02742E}.
Even
explicitly learning
tensorial properties like polarizability is not beyond the reach of ML based methods
through the introduction of tensorial neural networks 
or E(3)-equivariant neural networks
and have been demonstrated to yield good results for
the protonated water dimer~\cite{Zhang_10.1021/acs.jpcb.0c06926}
and for bulk liquid water~\cite{Schienbein_arXiv/2207.08661}, respectively. 
%
These seminal contributions highlight that machine learning 
is emerging as a promising route for the computation of 
quasi-exact
IR~spectra
at finite temperatures. 
%
Even the prediction of such spectra at ``gold standard'' quantum chemical accuracy
seems within reach today, yet it has not been achieved so far for 
complex molecular systems.\\
%
Some of us have recently shown how the 
PES of complex reactive systems, such as protonated water clusters of increasing size, 
can be represented at CC~accuracy using NNPs~\cite{Schran_10.1021/acs.jctc.9b00805,Schran_10.1063/5.0035438}.
%
These clusters have long been of significant interest due to their unique structural properties\cite{Nagashima1986/10.1063/1.450172,Miyazaki2004/10.1126/science.1096037,Zwier2004/10.1126/science.1098129,Singh2006/10.1002/anie.200504159},
their rich dynamics\cite{Asmis2003/10.1126/science.1081634,Headrick_10.1126/science.1113094,Vendrell2007/10.1002/anie.200702201} and their use as model systems for proton transfer 
reactions in aqueous solutions\cite{Wolke_10.1126/science.aaf8425}.
The effect of proton transfer on the IR spectrum in particular has been the 
subject
of 
extended interest,
which has 
spurred
key advances in IR spectroscopy\cite{Schwarz1977/10.1063/1.434748,Okumura1986/10.1063/1.451079}.
For this generic class of large \mbox{H-bonded} systems
we push the neural network approach 
to the next level 
by representing also dipole moments at 
close to
converged CC~accuracy.
In particular, we devise a very accurate 
neural network representation of the dipole moment surface (NN-DMS)
for the same set of protonated water clusters for which
we parameterized earlier a NN-PES
at essentially converged CC~accuracy~\cite{Schran_10.1021/acs.jctc.9b00805}. 
%
These developments enable 
the predictive calculation of IR~spectra 
at full CC~accuracy,
i.e. including the PES as well as the DMS, 
based on molecular dynamics simulations 
to take into account 
anharmonicity and finite-temperature effects~-- either
using classical point particles or quantum nuclei via path integrals. 
%
%
We note in passing that a
first application of our NN-DMS to most accurately compute the IR~spectrum 
of the bare Zundel cation (a.k.a. the protonated water dimer or \cf{H5O2+})
based on quasi-exact quantum dynamics propagation, thus
fully including nuclear quantum effects, 
has recently been 
published~\cite{Larsson_arXiv/2206.12029}.
%
%
As exemplified here, using differently sized protonated water clusters up
to the extended or solvated Zundel complex (a.k.a. the protonated water hexamer or  \cf{H13O6+})
our approach to accurate NN-DMSs holds great promise
for the predictive 
computation of vibrational spectra
in the future
toward deciphering the complex intra- and intermolecular couplings
within 
large molecular systems
on par with modern experimental spectroscopy.


\section{Neural Network Representation of Dipole Moments}
\label{sec:methodology}
%
In order to compute IR spectra one first needs an accurate description
of the electric dipole moment.
We present here an approach to describe the total electric dipole of 
molecular systems that is based on the high-dimensional neural network 
representation initially developed by Behler and 
Parrinello\cite{Behler_10.1103/PhysRevLett.98.146401} for the description of 
potential energy surfaces 
as
nowadays routinely used to develop 
high-quality machine learning potentials\cite{Behler_10.1002/anie.201703114}. 
Within the original approach, 
the total energy of a system composed of $N$~atoms is 
computed as a sum over atomic energies $\epsilon_i$, $E=\sum_{i=1}^{N}\epsilon_i$. 
These atomic energies are the output of associated element-based neural networks
that are fully connected feed-forward networks which take as input a set of 
descriptors of the local atomic environment around the considered 
atom\cite{Behler_10.1103/PhysRevLett.98.146401}
--~we use here the standard atom-centered symmetry functions~\cite{Behler_10.1063/1.3553717}.
It has been demonstrated earlier~\cite{Gastegger_10.1039/c7sc02267k}
that this approach can be generalized to the description of 
the total electric dipole moment $\vec{\mu}$ of molecular systems.
In this case, the electric dipole is computed through the standard expression 
for a system of $N$~classical point charges
\begin{equation}
   \vec{\mu} = \sum_{i=1}^N q_i\vec{r}_i,
   \label{eq:dip}
\end{equation}
where $q_i$ is the atomic charge associated with atom~$i$ and $\vec{r}_i$ 
its position.
The output of the element-based neural networks are now these atomic charges 
which thus depend on the local atomic environment.
\begin{figure}
    \centering
    \includegraphics[width=\linewidth]{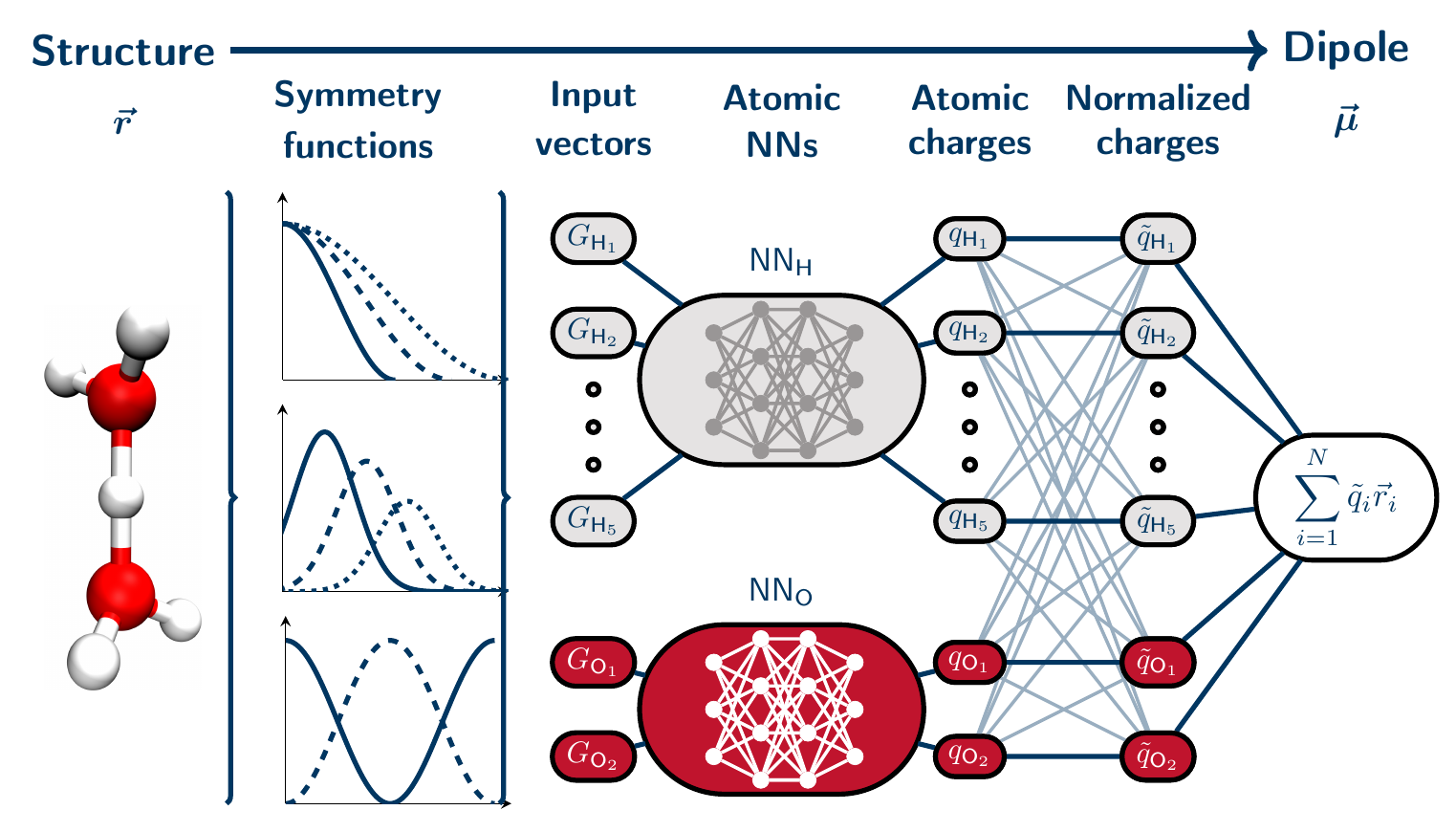}
    \caption{\textbf{
Schematic overview of the representation of the total electric dipole moment
by high-dimensional neural networks.}
In an initial step, the molecular structure is transformed by atom-centered
symmetry functions into atomic vectors.
Next, this atomic fingerprint is used as input for atomic neural networks
that output environment-dependent 
atomic
partial charges.
These charges are 
subsequently normalized 
in order to 
conserve
the total charge of the system
before computing the total electric dipole moment through the standard expression for
point charges.
}
    \label{fig:hd-nndms}
\end{figure}

This approach, schematically summarized in Fig.~\ref{fig:hd-nndms}, has been 
implemented in \texttt{RubNNet4MD}~\cite{RubNNet4MD-v2-dm}, our software package for training 
high-dimensional neural networks.
The neural network representation of the dipole moment surface (NN-DMS)
is trained to reproduce a set of reference dipole moments computed for 
various representative configurations of the system.
The cost function minimized during training is given by the root mean square 
error (RMSE) of the dipole moment components
\begin{equation}
   \mathcal{C}=\frac{1}{3M}\sum_{i=1}^{M}\sum_{\alpha=1}^3 (\mu^\text{NN}_{i,\alpha}-\mu^\text{ref}_{i,\alpha})^2
\end{equation}
with $\alpha$ running over the three components of the vector and 
$M$~the number of configurations in the training set.
Note that this cost function does not impose any constraint on the 
atomic charges so that the total charge of the system is not fixed.
In order to impose
the correct total charge~$Q$ of the system, 
the predicted electric dipole is
computed by equation~(\ref{eq:dip}) 
using normalized atomic charges $\tilde{q}_i$ given by
\begin{equation}
   \tilde{q}_i = q_i + \frac{1}{N}\left( Q - \sum_{j=1}^N q_j \right).
\end{equation}
This particular charge renormalization scheme has been previously suggested and successfully applied in the
PhysNet software package\cite{Unke_10.1021/acs.jctc.9b00181}.
%
Note that all atomic charges presented in the following are those normalized 
charges.
This approach is different than in previous work~\cite{Gastegger_10.1039/c7sc02267k}
where the total charge was included in the cost function thus
keeping it only close to the target total charge.

Once an accurate representation of the dipole moment as well as 
of
the potential energy surface is available, high-quality IR spectra can be computed using MD 
simulations
\cite{Ivanov_10.1039/C3CP44523B}.
Indeed, within linear response theory, 
the infrared~(IR) spectrum, described here by the linear 
IR~absorption coefficient per unit length $\alpha(\omega)$, can be expressed as
\begin{equation}
   \alpha(\omega) = \frac{\pi\beta\omega^2}{3Vc\epsilon_0 n(\omega)}
                    \frac{1}{2\pi}\int_{-\infty}^{+\infty}
                    \text{e}^{-i\omega t}\langle\vec{\mu}(0)\cdot \vec{\mu}(t)\rangle \text{d}t,
   \label{eq:alpha}
\end{equation}
where $\vec{\mu}(t)$ is the total electric dipole at time $t$, $n(\omega)$ the 
refractive index, which is essentially unity in gas phase, $\epsilon_0$ the 
vacuum permittivity, $V$ the volume (taken to be unity for gas phase), 
$c$ the speed of light and $\beta=1/k_\text{B}T$.
The division by three in the prefactor comes from an isotropic average over 
the polarization directions of the radiation field.
The interested reader is 
for instance
referred
to 
Ref.~\citenum{Ivanov_10.1039/C3CP44523B} for more details on theoretical 
spectroscopy from 
MD~simulations. 
The IR~spectrum is thus directly proportional to the Fourier transform of the 
time autocorrelation function (ACF) of the 
total 
dipole moment, 
\begin{equation}
   C(t)=\langle\vec{\mu}(0)\cdot \vec{\mu}(t)\rangle.
   \label{eq:acf}
\end{equation}
%
Note that formula~(\ref{eq:alpha}),
where the brackets denote the Boltzmann-weighted
statistical (NVT)~average at temperature $T$,
is obtained from the classical limit of the Kubo-transformed formulation 
of the quantum time~ACF\cite{Ramirez2004/10.1063/1.1774986}.
%
%
Such 
ACFs
can be directly obtained from MD simulations,
thus fully accounting for anharmonicities, 
mode couplings
and finite temperature effects.
%

\section{Computational Details}
\label{sec:comp-det} 
\subsection{Neural Network Training}
\begin{figure}
\centering
\includegraphics{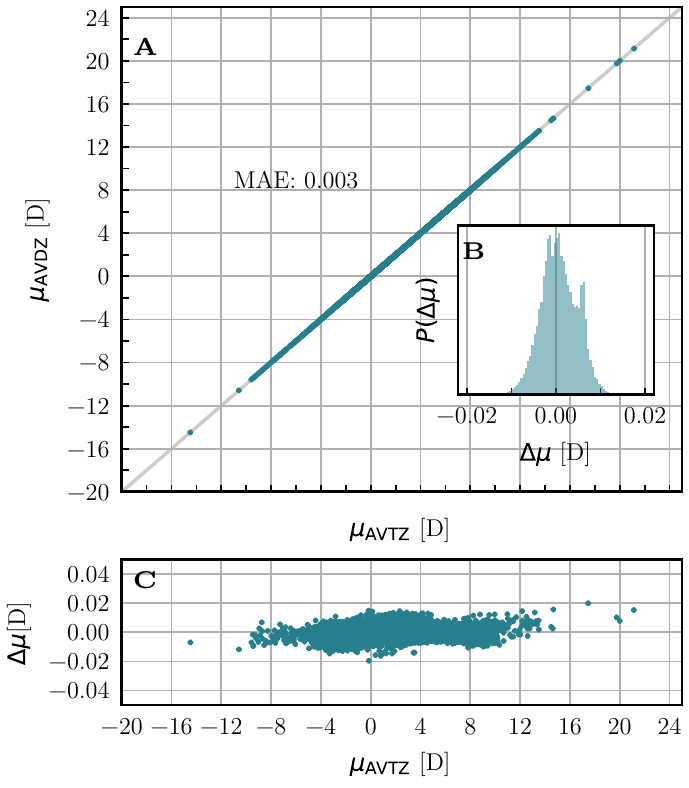}
\caption{\label{fig:AVDZ-validation}
   \textbf{
   Basis set convergence of coupled cluster dipole moments.
   }
   A:~Correlation between AVDZ and reference AVTZ dipole moment components.
   B:~Distribution of the respective differences $\Delta\mu$.
   C:~Differences of AVDZ versus AVTZ dipole moment components plotted against 
the corresponding
AVTZ reference dipole moment components.
   %
The dipole moment components $\mu_\alpha$, $\alpha = x, y, z$, rather than just its magnitude $|\vec{\mu}|$
and thus
the respective 
differences 
$\Delta\mu_\alpha$ are treated independently in all panels.
   Dipole calculations for 
   \num{10000} configurations of the Zundel cation 
   H$_5$O$_2^+$
   were carried out with the AVDZ and AVTZ basis sets
as specified in the text 
   to evaluate the degree of convergence achieved with the AVDZ basis set.
   The remaining parameters were kept identical between these two sets of calculations and are described in detail in
   the text.
   }
\end{figure}

%
The data set used to train the NN-DMS is composed of
the set of configurations 
generated previously by active learning to train the 
NN-PES~\cite{Schran_10.1021/acs.jctc.9b00805,Schran_10.1063/5.0035438}.
%
For each configuration, the reference electric dipole moments were computed with respect to
the origin of the coordinate system
using coupled cluster theory including singles, doubles and perturbative 
triple excitations in the density-fitting approximation DF-CCSD(T),
see Refs.~\citenum{Werner2018/10.1063/1.5020436} for background and methodology.
%
To remove any bias introduced
by translations,
all configurations were translated 
so that their
center of mass coincides
with the origin of the coordinate system.
Furthermore, each configuration was rotated at random
to remove rotational biases.
%
The augmented correlation-consistent basis set up to double zeta 
functions\cite{Kendall_10.1063/1.462569,Woon_10.1063/1.466439} (aug-cc-pVDZ or AVDZ) 
is used in combination with the explicitly correlated 
F12a~method\cite{Adler_10.1063/1.2817618,Knizia_10.1063/1.3054300} and
an adequate scaling of the triples~\cite{Knizia_10.1063/1.3054300},
thus providing the ``DF-CCSD(T*)-F12a/AVDZ'' approach (simply referred to as ``CC'' in this text).
%
The level of basis set convergence achieved when using that AVDZ basis 
was confirmed by reevaluating the dipole moments
using the computationally much more demanding larger AVTZ basis sets
for the approximately \num{10000} bare Zundel configurations, \cf{H5O2+}
in our training set
as shown in Fig.~\ref{fig:AVDZ-validation}.
%
The direct comparison of the components of all dipole moments at the two basis sets
yielded a negligible 
mean absolute error (MAE)
of \num{0.003}~D 
of AVDZ versus the reference AVTZ dipoles, 
which is roughly one
order of magnitude lower than the usual fitting error obtained with our NN-DMS.
In addition, the corresponding 
error distribution depicted in panel~B of Fig~\ref{fig:AVDZ-validation} reveals that
the errors rarely exceed 0.01~D.
Overall, inspection of the correlation between AVDZ and AVTZ results shows that
calculations with 
this
AVDZ basis set are well converged
given the purpose
and are, thus, suitable for use
as a 
reliable
reference 
to parameterize the NN-DMS. 

All these calculations have been carried out using the 
\texttt{Molpro} quantum chemistry package~\cite{molpro-review,MOLPRO-dipoles}.
Overall, the data set contains 54710 configurations of protonated water
clusters ranging from the protonated monomer \cf{H3O+}
up to the tetramer \cf{H9O5+} as well as \cf{H2O}.
A tenth of these points was chosen at random and
removed from the training set to serve as a test set.
The remaining points formed the training set to which the NN-DMS was fitted.
The NN 
to predict the dipole moment vectors $\vec{\mu}$ 
was constructed with 
two
hidden layers of 30~nodes each
using hyperbolic tangents activation functions.
An element-decoupled extended Kalman filter algorithm was used to optimize NN weights.
Underlying this work is the extension 
of our in-house \texttt{RubNNet4MD} package for generating high-dimensional neural networks
from energies (thus NN-PES) to vectorial properties, namely dipole moments
(thus NN-DMS)~\cite{RubNNet4MD-v2-dm}. 

\subsection{Infrared Spectra}
The same procedure has been applied
to generate IR spectra of both the bare Zundel cation (\cf{H5O2+}) 
and the much larger extended Zundel complex (\cf{H13O6+})
as follows: 
%
For each system, the canonical ensemble 
at temperature~$T$
was sampled by a single trajectory with timestep ${\Delta t=0.25}$\,fs using
a Langevin thermostat\cite{Grest_10.1103/PhysRevA.33.3628} with ${\tau=200}$\,fs. 
After equilibration 
for 10\,ps,
a phase space 
snapshot of this trajectory was taken every 500\,fs (2000~steps) and used to
spawn a non-thermostatted microcanonical (NVE) trajectory, which was then 
propagated for another 5\,ps with the same timestep. 
%
Translation of the molecule was removed in post-processing by moving the center of mass to
the origin of the coordinate system.
While not strictly necessary, this removes the spurious effects close to zero frequency 
introduced
by translational movement of a charged system.
The NN-DMS was applied to these centered trajectories to obtain the dipole moment ACF
and subsequently the IR spectrum,
see equations~(\ref{eq:acf}) and~(\ref{eq:alpha}).
For the Zundel cation, 60 trajectories were generated and processed in this manner.
%
At 300\,K, the larger 
\cf{H13O6+} complex 
can undergo 
thermal isomerizations
into conformations
different from that of the extended Zundel cation.
To obtain the IR~spectrum of only the 
genuine
extended Zundel cation,
trajectories exhibiting these rearrangements were not taken into account.
%
Instead, a larger number of NVE trajectories was generated 
and propagated for 5\, ps until 
60 
trajectories
had been obtained
which all correspond exclusively to the desired extended Zundel 
conformation of the protonated water hexamer \cf{H13O6+}. 

%
The Fourier transform
of the ACF obtained for each NVE~trajectory
has been performed by applying a
Hann window spanning the entire trajectory
to ensure minimal smoothing.
The reported IR~spectra at temperature~$T$ have been obtained
as the average over all 60~individual NVE simulations, spawned from the NVT ensemble
as described above.
%
The total dipole moment vectors of the protonated water clusters were autocorrelated
every 2\,fs in order to provide
the required resolution to numerically converge the reported IR~spectra up to 4500\,\icm.

The underlying molecular dynamics simulations with classical
nuclei have been carried out using the \texttt{CP2k} software package
\cite{cp2kwebsite,hutter_wires2014} on a previously 
parameterized and published NN-PES~\cite{Schran_10.1021/acs.jctc.9b00805}.
This NN-PES describes the Born-Oppenheimer 
energy landscape of protonated water clusters at
CCSD(T*)-F12a/AVTZ accuracy 
that is largely consistent with the 
DF-CCSD(T*)-F12a/AVDZ accuracy achieved here for the
dipole moment vector surface of the same protonated water clusters
by virtue of the present NN-DMS parameterization, both being close to the
complete basis set limit due to using the explicit correlation factor~F12a
in the reference CC~calculations. 
%

\section{Reaching Coupled Cluster Accuracy for Dipoles}
\label{sec:fitting}

%
\begin{figure}
\centering
\includegraphics{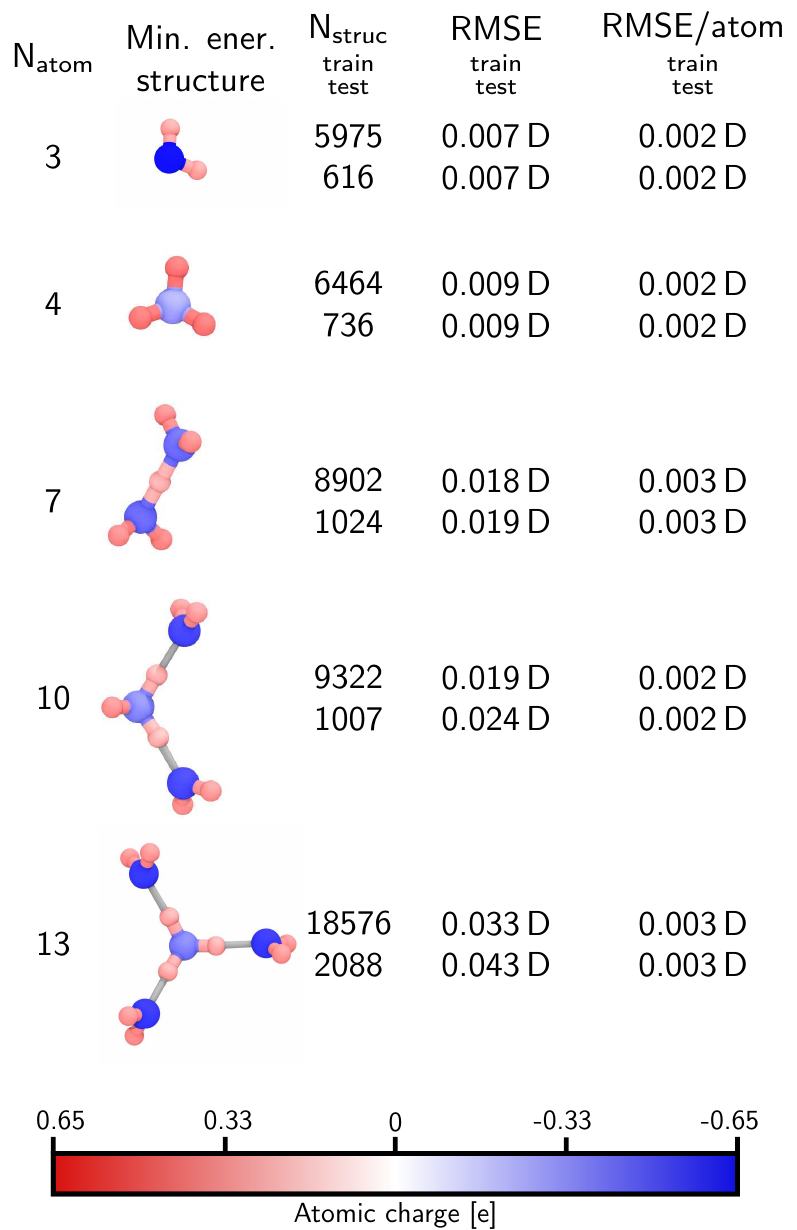}
\caption{\label{fig:struc-rmse}
\textbf{Overview over all systems used for the 
parameterization of
the NN-DMS for protonated water clusters up to the protonated water tetramer.}
The color of the atoms in the second column encodes the 
normalized partial charges $\tilde{q}_i$ of the atoms within each respective cluster 
in its global minimum energy configuration
as specified by the color scale at the bottom.
Note that these partial charges are exclusively fitted to reproduce each cluster's dipole
moment without any chosen bias. 
The third column gives the number of structures in the training and
test sets and the fourth one reports the root mean square error for each
system in the training and test sets.
The fifth column provides the root mean square error divided by the number
of atoms in the respective cluster.
}
\end{figure}

\begin{figure}
\centering
\includegraphics{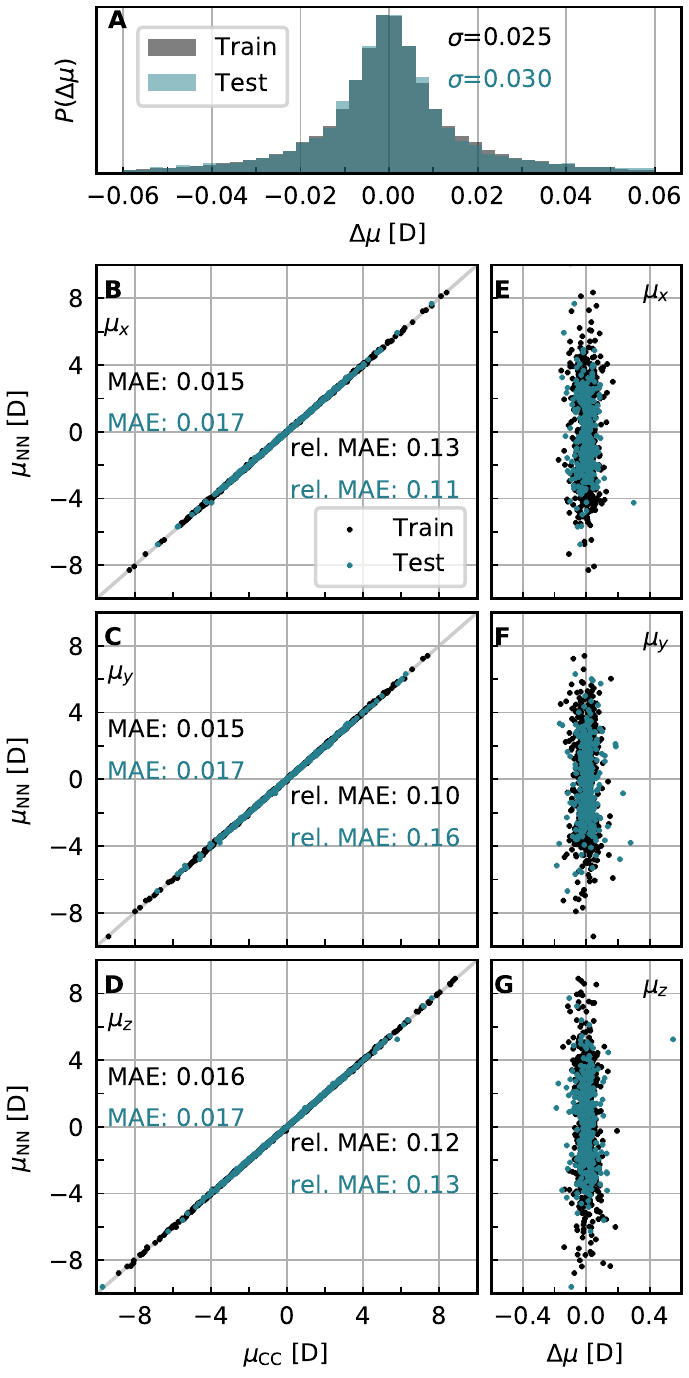}
\caption{\label{fig:dip_corr}
\textbf{Validation of the training of the NN-DMS.}
A:~Distribution of errors 
of the total dipole 
moment components
across training (black) and test (blue) set.
B~to~D:~Correlation of total dipoles from the NN-DMS with the reference CC~dipoles.
%
E~to~G:~Errors $\Delta\vec{\mu}=\vec{\mu}^\text{NN}-\vec{\mu}^\text{CC}$ 
as a function of the 
dipole moment values predicted by the NN-DMS.
The very few outliers 
in panels E and G correspond to 
test set configurations with 
highly distorted structures which are 
very high up in potential energy with respect to the relevant PES~minima,
thus leading to a very low density of training points in 
such ``rate event regions'' of 
configuration space.
%
The total dipole moment components $\mu_\alpha$, $\alpha = x, y, z$, rather than just its magnitude $|\vec{\mu}|$
and thus 
the respective errors $\Delta\mu_\alpha$ are treated independently in all panels. 
%
Both training and test sets contain 
all configurations of the protonated clusters from \cf{H3O+} to \cf{H9O4+} and \cf{H2O}.
}
\end{figure}

We use the high-dimensional neural networks approach
to develop a highly accurate
NN-DMS
for protonated water clusters.
This development explicitly includes clusters from
the hydronium ion \cf{H3O+} up to the tetramer \cf{H9O4+}, 
as well as the water molecule.
The model is trained against a set of reference dipoles that have
been computed using electronic structure calculation at 
the
coupled cluster level 
including single, double and perturbative triple excitations, CCSD(T), 
for a large number (54710) of representative 
configurations of the different clusters.
These atomic configurations have been optimally selected previously through 
an automated fitting procedure used to develop an accurate neural network potential 
energy surface (NN-PES) for protonated water clusters at the 
CCSD(T) accuracy 
level~\cite{Schran_10.1021/acs.jctc.9b00805}.
Additional computational details are provided in section~\ref{sec:comp-det}.

To ascertain the quality of the resulting NN-DMS, a number of stringent tests have been performed.
Our final NN-DMS predicts training and test configurations with RMSEs of
0.025 and 0.035\,D, respectively.
%
%
To compare the performance of the model 
for the differently sized clusters covered by the data set, 
we disentangle the total RMSE into the various clusters as shown in Fig.~\ref{fig:struc-rmse}.
It is clear that 
this NN-DMS
accurately describes the dipoles of all clusters
up to the tetramer.
This analysis reveals that the fitting error increases
slightly with cluster size.
In particular, \cf{H2O} and \cf{H3O+}
exhibit extremely low RMSE values below 0.01\,D for both
training and test sets.
This does not come as a surprise since the larger number
of degrees of freedom make the representation of the reference dipole moment 
more complex for larger clusters.
Nevertheless, we observe a very accurate 
overall representation of the dipole moments with RMSE values well below $0.05$\,D.
To account for the 
additional complexity
caused by the increased number of degrees of freedom,
we also report the error divided by the number of atoms in each cluster in Fig.~\ref{fig:struc-rmse}.
This reveals a constant RMSE of $0.002$ to $0.003$\,D per atom for all clusters,
indicating that the good performance of the NN-DMS is retained for
all clusters.
%
Our errors compare well
with Ref.~\citenum{Gastegger_10.1039/c7sc02267k} 
in which a 
MAE
of $0.016$~D is reported for methanol over a range of $0.723$~D.
Here we obtain a total MAE of $0.015$~D for 
five different species
over a range of about $16$~D
and a MAE of $0.011$\,D over a range of $8$\,D for \cf{H5O2+},
the cluster closest to methanol in terms of size.
%
The obtained accuracy is particularly impressive,
given the diverse nature of our data set including not only
differently sized clusters, but also various isomers 
and reactive rearrangement corresponding to proton transfer
\cite{Schran_10.1021/acs.jctc.9b00805}.
%

It has previously been shown that the environment dependent atomic charges
output by the neural network can
capture some aspects of the chemistry underlying a system~\cite{Gastegger_10.1007/978-3-030-40245-7_12}.
We verify this effect here by studying 
the predicted atomic charges for each cluster in its global minimum energy structure,
as represented in Fig.~\ref{fig:struc-rmse} where each atom
is colored according to its 
partial
charge as assigned by the NN-DMS.
As expected, the predicted charges correlate strongly with chemical intuition:
The more electronegative oxygens are always associated with a negative atomic charge 
while the hydrogens are then assigned a positive charge.
Moreover, for the trimer and tetramer, the central oxygen 
is considerably less negative than the outer 
oxygen atoms, as expected for a hydronium-like core
with three hydrogen atoms.
It should be kept in mind that such atomic charges,
although resembling other charge partitioning
schemes, remain parameters of the model and are
not uniquely determined.
Nevertheless, they can prove useful for qualitative
analysis of related properties, such as the 
electrostatic potential as shown previously~\cite{Gastegger_10.1007/978-3-030-40245-7_12}.
%
Finally, we analyze the performance of the model
in more detail by looking at the correlation
of the predicted dipole components with 
the reference ones as shown in Fig.~\ref{fig:dip_corr}.
Panel~A of this figure shows a histogram of the prediction errors, in which each
component $\mu_\alpha$, $\alpha = x, y, z$ of the total dipole moment $\vec{\mu}$ 
was treated as an independent prediction, i.e. each configuration
contributes three 
entries
to this 
(and the related) histograms. 
Histogram~A is a simple 
yet useful first
indicator 
to support
the overall quality of the model.
The distributions of both, training and test data 
are very narrow with a standard deviation of 0.025\,D and 0.03\,D, respectively.
%
With the histogram tails trailing off 
already
around 0.04\,D, the overwhelming majority of 
points have very small errors and even the worst predictions are still satisfactory.
The bottom three rows of Fig.~\ref{fig:dip_corr} show a point-by-point comparison between
CC and NN~dipoles separated into $x$-, $y$- and $z$-component
of the total dipole moments.
On the left, the CC~dipole component is plotted against the NN~dipole component, 
whereas the right-hand 
panels
show the NN~dipole component against the error,
\mbox{$\Delta{\mu}_\alpha={\mu}_\alpha^\text{CC}-{\mu}_\alpha^\text{NN}$,} thus 
providing a detailed view on outliers and potential systematic errors.
Overall, essentially perfect correlation between the
prediction and the reference is observed.
More importantly there is almost complete absence
of outliers or satellite groups.
This corroborates that the 
NN-DMS
is providing convincing
accuracy for all considered clusters.
This analysis does not reveal any differences between
the three 
spatial
dimensions, as
required for a rotationally invariant representation.
Furthermore, it can be seen that test points are
represented at essentially the same accuracy as the training points.
Overall, all 
these tests paint the picture of a highly accurate 
NN~representation of the full dipole moment surface of the five different molecular species
for any given configuration considered in the data set.

\begin{figure*}
\centering
\includegraphics[width=\textwidth]{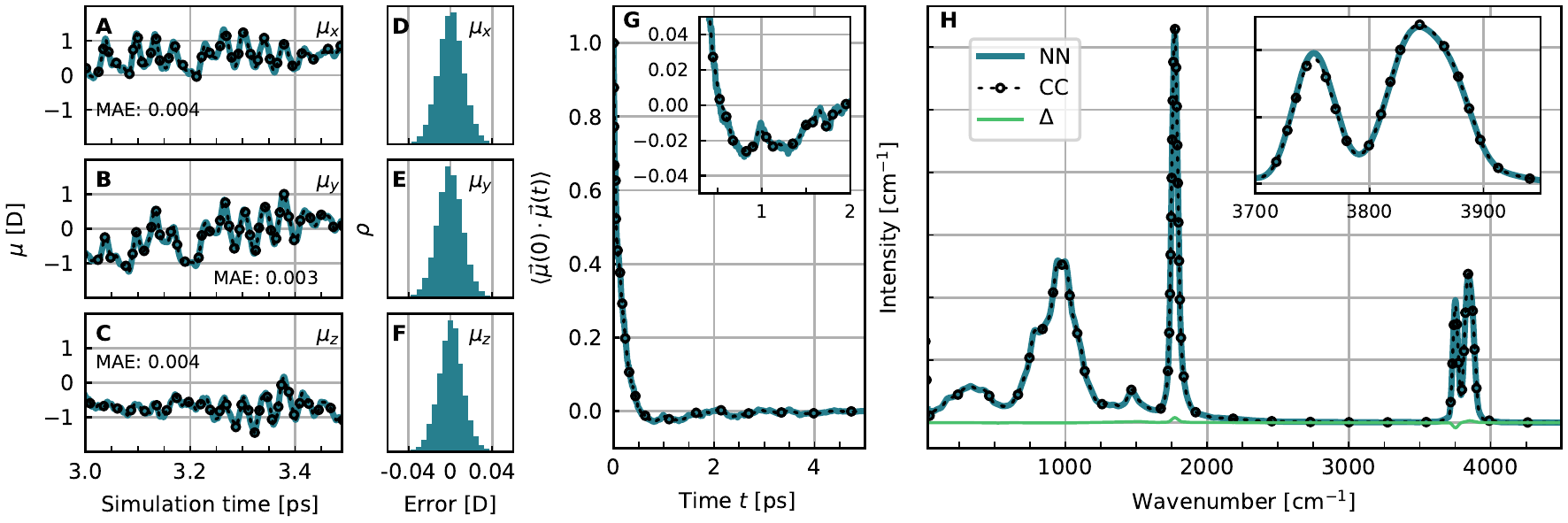}
\caption{\label{fig:w2h+-spectra}
\textbf{Performance of the NN-DMS for the bare Zundel complex, \cf{H5O2+}.}
%
A to C: Short excerpts from MD trajectories at 100\,K comparing 
the dipole moments from explicit CCSD(T)~calculations to those from the NN-DMS separated into their $x$-, $y$- and $z$-components; 
the corresponding MAE is provided in the respective panels. 
%
D to F: Corresponding error distributions over all 
60~NVE trajectories generated to compute the IR~spectrum in panel~H. 
%
G: Dipole autocorrelation function from NN and CC~dipole moments.
%
H: IR~spectrum of the protonated water dimer at 100\,K computed from
NN and CC~dipole moments from the same set of trajectories.
%
The magnified peak in the inset visualizes some very small difference 
in peak intensity whereas the $\Delta$~line quantifies 
the spectral differences as a function of frequency.
}
\end{figure*}

\section{Validation on Dipole Time Evolution and IR Spectra}
\label{sec:advanced}
%
So far, the analysis of the performance of the NN-DMS
has concentrated on the fitting accuracy compared
to the reference method.
In order to put the methodology to a much more
stringent test, we 
focus
here on the ability of our dipole moment surface  
to reproduce dipole fluctuations and 
real-time
dynamics as 
generated
in the course of 
realistsic molecular dynamics simulations. 
%
This is a 
tough challenge to the NN-DMS
since the 
typically subtle time-dependent 
changes of the dipole moment
during molecular vibrations is what gives rise to the IR response.
%
Hence, even tiny
errors can 
have a large impact on the accuracy of the computed IR~spectra.
%

We carried out the required benchmarking for
the smallest protonated water cluster featuring a shared proton,
for which the 
very demanding
CCSD(T)~reference dipoles can still be obtained for 
the required large number of configurations 
of about~\num{300000} in total. 
To do so, we computed the dipoles of the 
bare 
Zundel cation, \cf{H5O2+},
using both the CC~reference method and the NN-DMS
over the whole length of the molecular dynamics trajectories;
recall that 60~independent trajectories have been computed
to rigorously converge the IR~spectrum. 
This is of course a very expensive test since many CC calculations 
are required, but it 
grants
access to the computation of the 
converged 
ACF
and thus the IR spectrum from both, 
the reference CC method and the NN-DMS allowing us to perform 
a one-to-one
comparison. 
The results are displayed in Fig.~\ref{fig:w2h+-spectra}, 
where panels~A to C show a 
tiny yet
representative excerpt from 
a random single trajectory out of 60~in total 
used to 
compute 
the IR spectra, separated into the $x$-, $y$- and $z$-components
of the total dipole moment $\vec{\mu}$. 
It can be seen clearly that the 
physical
fluctuations caused by configurational changes exceed by far any errors
introduced by the NN-DMS.
Moreover, the 
histograms in panels~D to~F 
quantify the errors based on all trajectories 
proving that the 
NN-DMS
retains its excellent performance also in the course of 
the extensive
molecular dynamics simulations.
%

To 
compute the
IR~spectrum from these accurate dipole moments, the 
ACF of the total dipole moment 
needs to be computed.
Panel~G of Fig.~\ref{fig:w2h+-spectra} depicts this 
dipole autocorrelation function
as computed using the NN-DMS 
in direct comparison to
the reference CCSD(T) dipoles.
It can be seen that the 
ACF
retains the same quality already observed in the dipole moment trajectories.
Finally, as an end-to-end validation, we have computed 
the IR spectrum of \cf{H5O2+} from the dipole~ACF
in an effort to directly compare the 
NN-DMS performance to explicit CCSD(T)~calculations for 
a most sensitive observable that is experimentally accessible. 
Comparing the spectra shown in panel~H of Fig.~\ref{fig:w2h+-spectra} 
reveals almost perfect agreement between CC and NN spectra.
Only two regions of the spectrum show small deviations, 
one of which is the wavenumber range between 3700 and 3800\icm, 
which needs to be magnified (see inset in~H) to visualize them. 
The other region exhibiting similar errors is the peak at
about
1850\icm.
%
In both cases, however, the error is negligible on the 
intensity 
scale of the spectrum,
whereas no effects on the peak positions and their shape can be detected. 
This final test 
demonstrates
that the NN-DMS indeed 
allows one to rigorously compute 
IR~spectra at essentially converged CC~accuracy.

%
Finally, we stress that the aim of computing the IR spectrum of \cf{H5O2+} 
as depicted in panel~H of Fig.~\ref{fig:w2h+-spectra}
was to exclusively test the NN-DMS while using classical~MD trajectories 
at 100~K to generate the necessary trajectories
(since that computationally economic approach allowed us to 
explicitly re-compute the
dipole moments directly using the CCSD(T) reference method
for the same set of sampled configurations). 
%
To this end, we employ our recently published highly accurate NN-PES 
to describe protonated water clusters 
\cite{Schran_10.1021/acs.jctc.9b00805}, with an RMSE of
0.06\,kJ/mol per atom while spanning an
energy range of several 100\,kJ/mol.
%
We refer the interested reader to the original publication for detailed
benchmarks of proton transfer paths and MD~sampling using this NN-PES,
and in particular to Figure~7 therein where we thoroughly demonstrated 
the accuracy of this NN-PES by comparing to 
coupled cluster single-point energies that have been computed
explicitly along representative trajectory segments 
generated at several temperatures. 
%
%
%
%
%
%
Clearly, classical~MD simulations are unable to correctly
describe the structural dynamics of \cf{H5O2+} owing to
the extremely complex quantum dynamics of the 
Zundel cation~\cite{Vendrell_10.1002/anie.200702201}.
%
Here, we refer to most recent progress~\cite{Larsson_arXiv/2206.12029} on 
accurate quantum dynamics and IR~spectroscopy of \cf{H5O2+}
that results from using the present NN-DMS 
for protonated water clusters in conjunction with the
the existing NN-PES~\cite{Schran_10.1021/acs.jctc.9b00805}
both a essentially converged coupled cluster accuracy. 

\begin{figure*}
\centering
\includegraphics[width=\textwidth]{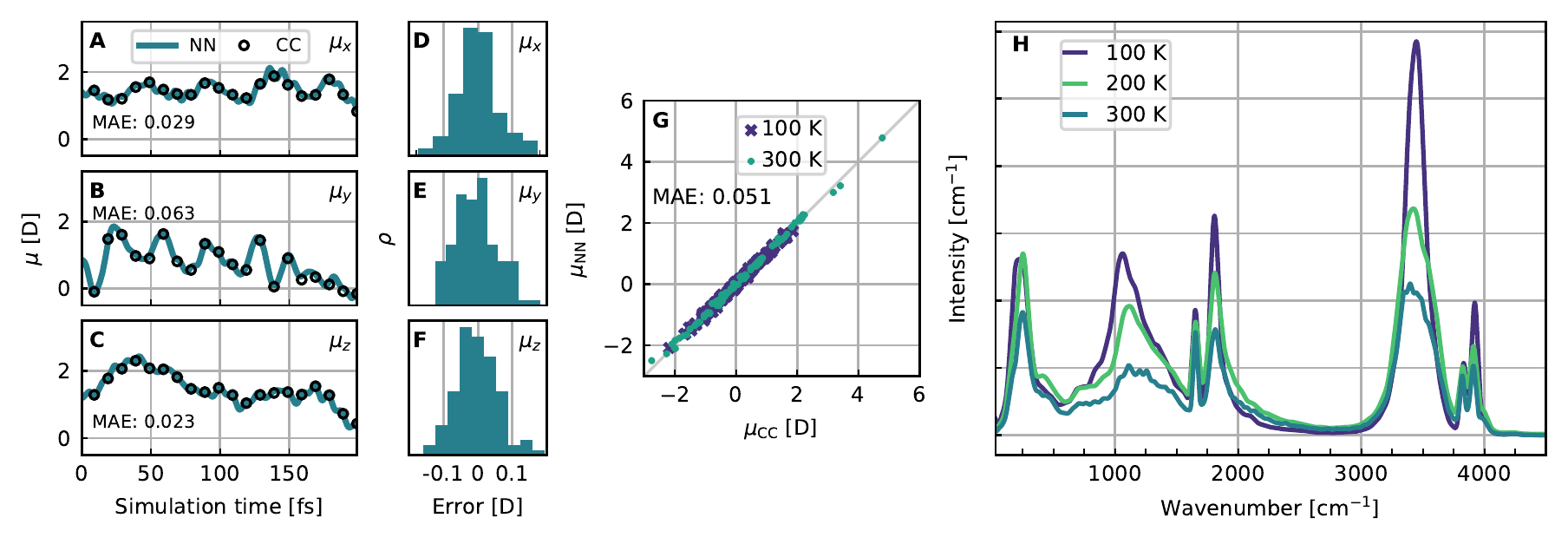}
\caption{\label{fig:w6h+-validation}
\textbf{Transferability of the NN-DMS for the extended Zundel complex, \cf{H13O6+}.}
%
A to C: Short excerpts from MD trajectories at 100\,K comparing 
the dipole moments from explicit CCSD(T)~calculations to those from the NN-DMS separated into their $x$-, $y$- and $z$-components;
the corresponding MAE is provided in the respective panels.
%
D to F: Distribution of errors from the short dipole trajectory excerpt
(see panels A to C) and 120 additional configurations randomly extracted 
from MD trajectories at 100 and 300\,K (see panel G).
%
G: Correlation of NN dipoles and reference CC dipoles
taken from the small MD trajectory excerpt as well as
additional configurations randomly extracted from the MD trajectories
used to generate the spectra. Associated error distribution 
are given in panels D to F.
%
H: IR~spectrum of the 
extended Zundel cation at 100\,K, 200\,K and 300\,K,
computed from NN~dipoles based on 60~NVE trajectories at each temperature. 
}
\end{figure*}

\section{Prediction of IR spectra of larger clusters}
\label{sec:extrapolation}

After the assessment of the quality of the NN-DMS
for clusters within the training data, we finally push the model to the limit and 
test our model for 
species
not considered in the training process~--
thus probing transferability by entering the extrapolation regime of the dipole network.
It has previously been shown that the NN-PES trained on
these clusters up to the protonated tetramer 
retained its predictive power also for larger clusters such as
the protonated water hexamer, despite not being part of the training data
\cite{Schran_10.1063/5.0035438}.
%
This can be attributed to the similarity of the larger clusters
to the chemical space spanned by the training set.
In the following we will show that this transferability to more complex
situations is also 
achieved by
the NN-based dipole surface by
performing explicit 
calculations for the protonated water hexamer in its 
extended Zundel cation 
conformation. 
These advanced simulations reveal
that our ML approach 
provides
access to 
observables such as
the IR~response at essentially converged CC~accuracy of highly complex systems
hitherto accessible at such level of theory
beyond the systems considered during training of the model.

As a matter of fact,
converged CCSD(T) 
calculations of the protonated water hexamer are prohibitively
expensive
considering the number of single-point dipole calculations required to
statistically converge its finite-temperature IR~spectrum, 
thus
preventing us from performing the 
trajectory-based NN~to~CC comparison that was still computationally feasible 
for the protonated water dimer.
Instead, we 
invoke alternative benchmarks as follows. 
%
First, we 
assess
the performance of the NN-DMS 
to describe the real-time dynamics of the dipole moment
over~200\,fs of a representative MD~simulation 
with respect to the explicit CCSD(T) reference.
%
The evolution of dipole moments along 
this piece taken from a random trajectory out of 60~used to compute the final IR~spectrum 
at $300$\,K is shown in panels~A to C of Fig.~\ref{fig:w6h+-validation},
overlaid by explicitly computed CCSD(T) dipole moments every $10$~fs.
%
Similar to what was observed for the bare Zundel cation, the
NN-DMS prediction agrees very well with the CC reference
and, most importantly, captures the subtle but crucial dipole dynamics which fully determines the IR~spectrum. 
%
Thus, this NN-DMS is transferable to more complex species than those
used to learn it. 
Secondly, we analyze the long-time stability 
as to 
the predictive power of the NN-DMS by 
quantifying
the correlation of NN~predictions and the
exact CCSD(T)~dipole moments as obtained from explicit single-point CC~calculations for a 
realistic 
validation set build from 
the extensive MD simulations used to converge the IR~spectrum of the extended Zundel complex,
namely 120~configurations 
randomly
extracted from 
all 60~NVE trajectories. 
%
%
The correlation analysis in panel~G of Fig.~\ref{fig:w6h+-validation} demonstrates
that also the large dipole fluctuations as observed only during extensive
MD trajectories
are well reproduced, thus supporting further the transferability of the NN-DMS. 
Although the number of points is relatively small, 
owing 
to the tremendous cost of explicit CCSD(T)
dipole moment calculations for the extended Zundel complex, it can already 
be seen from that plot that the NN computations closely match the CC calculations.
The detailed error analyses in terms of histograms depicted in panels~D to~F
underlines this assessment by revealing essentially no errors above
0.1\,D, being only slightly larger than in the interpolation regime.
Despite this expected 
small decrease in accuracy, the NN-DMS is yielding
convincing results across the board
as can be seen from the small RMSE of $0.065$\,D that we obtained
within this validation step. 

%

As a final test for the transferability of the NN-DMS,
we decided to compute the 
full IR~response of the extended Zundel complex.
%
The simulated IR~spectra 
at 100\,K, 200\,K and 300\,K are shown in
panel~H of Fig.~\ref{fig:w6h+-validation}.
The most prominent feature of these spectra is the broad signal
from about
3300 to 3600\icm{} which
can be ascribed to the red-shifted 
O-H stretch vibrations of
those
hydrogen atoms 
which are 
involved in intermolecular H-bonds within the complex. 
%
%
These are clearly distinguishable from the high-frequency vibrations of
the outermost hydrogen atoms which do not form hydrogen bonds 
and yield two very sharp signals at 3800 to 4000\icm
as is well-known from such largely unperturbed free or dangling OH~bonds. 
%
%
At 1600 to 1800\icm, another prominent pair of peaks is visible, 
corresponding to 
%
frequencies known from coupled 
shared-proton and water bending motion
in the bare Zundel complex. 
At even lower frequencies, a 
broad yet overall intense
signal can be observed between 800 and 1500\icm, 
%
%
%
which covers the frequency window where the proton transfer doublet 
within the bare Zundel complex is located. 
The spectra at 200\,K and 300\,K spectrum show
the expected thermal broadening with regard to the
spectrum at 100\,K, but otherwise retains the same features. 
%

It is reassuring 
to conclude 
that the finite-temperature IR~spectra we computed 
for \cf{H13O6+}
from the highly accurate NN-DMS 
agrees in its features with the IR~response expected 
for the protonated water hexamer in the extended Zundel conformation.
%
This corroborates that the NN-DMS is able to accurately
predict the IR~response even for
larger systems, not considered during the training of 
that property surface. 
This opens up the possibility to systematically push the limits of CC~theory
to enable the 
quantitatively predictive computation of physical observables such as 
anharmonic IR~spectra at finite temperatures, otherwise inaccessible
by explicit electronic structure calculations.
%

\section{Conclusions and Outlook}
\label{sec:con}

In summary, we have outlined a clear strategy for the prediction of 
anharmonic finite-temperature 
IR~spectra
at coupled cluster accuracy using neural network representations.
As demonstrated for differently sized protonated water clusters,
learning 
both,
the potential energy and dipole moment surfaces by readily
developed machine learning models enables the computation of the 
converged 
IR response through 
exhaustive 
molecular dynamics simulations.
In addition, we have shown that such models can be applied to larger
molecular species, 
not considered in the development of the model, if sufficient care is taken in 
learning the model and validating its predictions.
This makes it possible to systematically push the limits of 
``gold standard''
coupled cluster theory to
larger system sizes in order to enable the prediction of 
physical properties such as 
IR~spectra that are 
otherwise inaccessible 
in view of the enormous computational cost
of explicit CCSD(T) calculations. 
%

While the IR spectra shown here are based on a classical
description of the nuclei, it is known that nuclear quantum effects
can have a significant impact on dynamical properties --- in particular
for high frequency modes.
Given that nuclear quantum effects have been accounted for during the
development of the training set~\cite{Schran_10.1021/acs.jctc.9b00805},
we plan to systematically explore 
as the next logical step
approximate quantum dynamics
approaches, such as centroid and ring polymer molecular 
dynamics~\cite{Craig_10.1063/1.1777575,Cao_10.1063/1.467176,Althorpe_10.1140/epjb/s10051-021-00155-2}
for the prediction of IR spectra using
the same
machine learning approach.
Other avenues to accurately incorporate the quantum dynamics of the nuclei
could be quasi-classical approaches~\cite{Makri_10.1146/annurev.physchem.50.1.167,%
Beutier_10.1080/00268976.2015.1064550,Micciarelli_10.1063/1.5096968}
or 
multi-configuration
time-dependent Hartree
(MCTDH) 
methodologies~\cite{%
Meyer_10.1016/0009-2614(90)87014-I,Manthe_10.1063/1.471847}
as previously successfully applied 
to the bare Zundel cation, \cf{H5O2+}~\cite{%
Vendrell_10.1002/anie.200702201,
Vendrell_10.1002/anie.200804646}.
%
Indeed, our new NN-DMS has been used most recently 
in conjunction with our NN-PES~\cite{Schran_10.1021/acs.jctc.9b00805}
for protonated water clusters 
to generate quasi-exact quantum dynamics of the bare Zundel cation
including the corresponding highly accurate 
IR~spectrum~\cite{Larsson_arXiv/2206.12029}
in excellent agreement with experiment based on
advanced vibrational tree tensor network states (TTNS)
techniques~\cite{Larsson_10.1063/1.5130390}, 
thus going beyond multilayer~MCTDH methods. 
%
Given the convincing performance 
of our NN-DMS 
with respect to 
explicit coupled cluster 
calculations, 
we think that our machine learning approach holds great promise
for the predictive 
computation of 
physical observables such as 
vibrational spectra, 
notably including 
the detailed understanding of the intra- and intermolecular
couplings within molecular systems of increasing complexity
way beyond that of the protonated water dimer. 
%

\section*{Data Availability}
\label{sec:data} 
The data used to train and validate the 
NN-DMS
presented in this paper 
together with the final NN-DMS paremeterization 
are available
in the Supporting Information to this publication 
(data-set: All coupled cluster dipole training data 
of the NNP model sorted by cluster size;
benchmarks: All benchmark coupled cluster calculations to 
compute the spectra of \cf{H5O2+} and 
120~reference dipole moments of \cf{H13O6+};
example-molpro-input: An exemplary \texttt{Molpro} input
file for the computation of the coupled cluster dipole moments;
model: All parameters of the NN-DMS) 
and are also 
publically accessible via
\texttt{nn-dms-supporting-data-v1.0.zip} 
at \texttt{https://doi.org/10.5281/zenodo.6901468}.
%

\section*{Code Availability}
\label{sec:code} 
All CC~calculations were performed using the \texttt{Molpro}
quantum chemistry package~\cite{molpro-review,MOLPRO-dipoles}.
The 
NN-DMS
has been constructed using the \texttt{RubNNet4MD} 
package~\cite{RubNNet4MD-v2-dm},
while the NN-PES used is available within the Supporting Information of
Ref.~\citenum{Schran_10.1021/acs.jctc.9b00805}.

\begin{acknowledgments}
%
We 
would 
like to thank 
Harald Forbert (Bochum) not only for insightful discussions but for providing
us with a program to efficiently and accurately compute IR~spectra
from MD~simulations as well as
Hans-Joachim Werner (Stuttgart) for help with the \texttt{Molpro} program package 
R.B. acknowledges funding from the
\textit{Studienstiftung des deutschen Volkes}
and
C.S. acknowledges partial financial support from the
\textit{Alexander von Humboldt-Stiftung}.
%
%
Funded by the \textit{Deutsche Forschungsgemeinschaft} 
(DFG, German Research Foundation) under Germany's
Excellence Strategy~-- EXC~2033~-- 390677874~-- RESOLV
as well as by the individual DFG~grant \mbox{MA~1547/19} to D.M.
%
This work is
supported by the ``Center for Solvation Science ZEMOS''
funded by the German Federal Ministry of Education and Research
and by the Ministry of Culture and Research of North Rhine-Westphalia. 
%
The computational resources were provided by
HPC@ZEMOS, HPC-RESOLV, and BoViLab@RUB.
\end{acknowledgments}

\section*{References}

\bibliographystyle{achemso}
\bibliography{bibliography}

\newpage
\begin{figure}
    \centering
    \includegraphics[width=\linewidth]{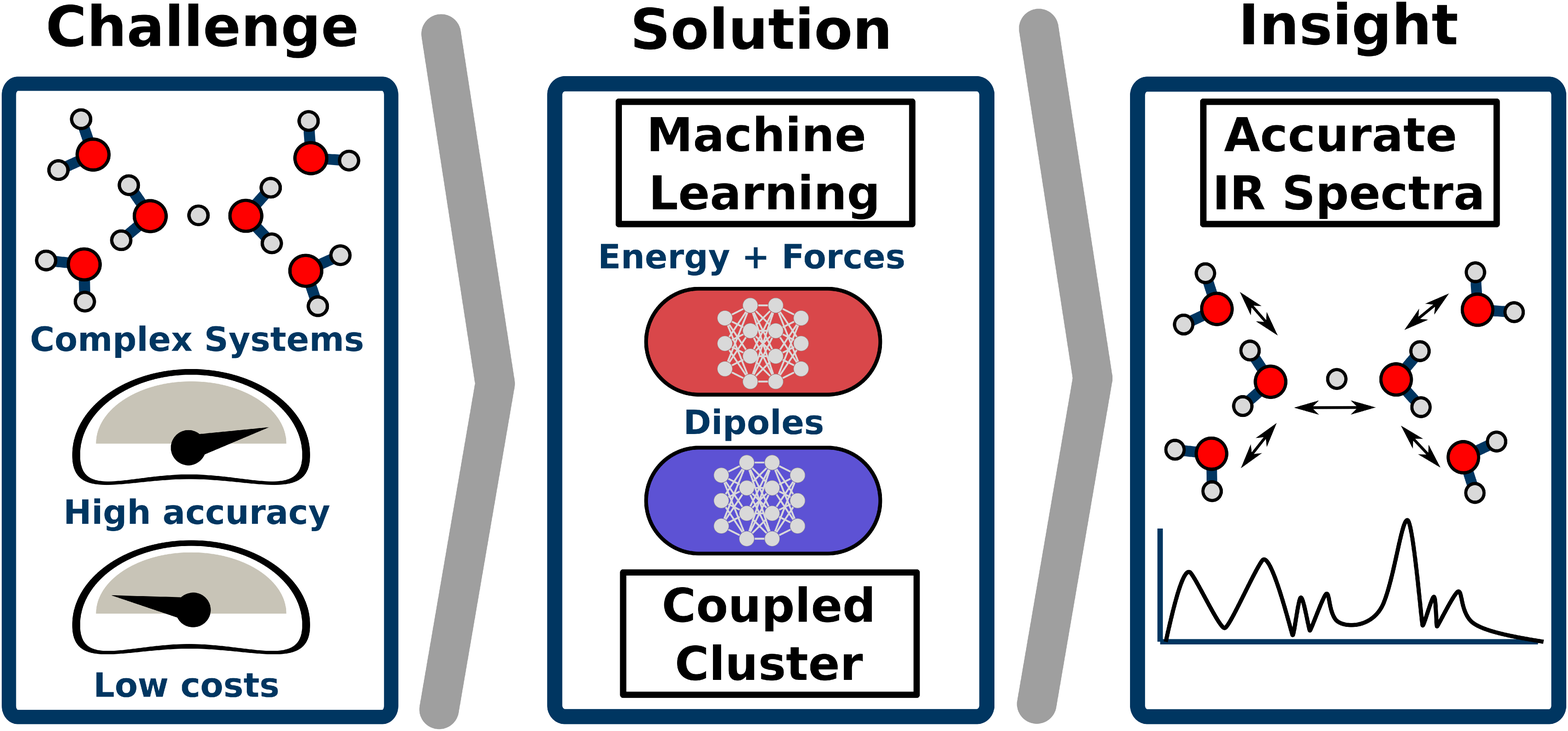}
    \caption{For Table of Contents Only}
    \label{For Table of Contents Only}
\end{figure}

\end{document}